\documentclass[12pt]{article} 

\begin{document}

\newcommand{\sect}[1]{\setcounter{equation}{0}\section{#1}}
\renewcommand{\theequation}{\thesection.\arabic{equation}}
\newcommand{\be}{\begin{equation}}
\newcommand{\ee}{\end{equation}}
\newcommand{\bea}{\begin{eqnarray}}
\newcommand{\eea}{\end{eqnarray}}
\newcommand{\beano}{\begin{eqnarray*}}
\newcommand{\eeano}{\end{eqnarray*}}
\newcommand{\rhotimes}{\mbox{\raisebox{-1.2ex}{$\stackrel{\displaystyle\otimes}
{\mbox{\scriptsize{$\rho$}}}$}}}
\newcommand{\Tt}{{\widetilde{T}}}
\newcommand{\wT}{{\widetilde{T}}}
\newcommand{\Qt}{{\widetilde{Q}}}
\newcommand{\hI}{{\hat{I}}}
\newcommand{\pv}{{\mbox{pv}}}
\newcommand{\Ygl}{{$Y(gl(N))$}}
\newcommand{\Ysl}{{$Y(sl(N))$}}
\newcommand{\calr}{\mbox{${\cal R}$}}
\newcommand{\cS}{\mbox{${\cal S}$}}
\newcommand{\cB}{\mbox{${\cal B}$}}
\newcommand{\cC}{\mbox{${\cal C}$}}
\newcommand{\cZ}{\mbox{${\cal Z}$}}
\newcommand{\bin}[2]{{\left( {#1 \atop #2} \right)}}

\pagestyle{empty}
\rightline{May 1999}

\vfill

\begin{center}
{\LARGE\bf Yangian symmetry in the
Nonlinear Schr{\"o}dinger hierarchy}
\\[2.1em]
\end{center}

{\Large
M. Mintchev$^{a}$\footnote{mintchev@difi.unipi.it},
E. Ragoucy$^{b}$\footnote{ragoucy@lapp.in2p3.fr}, 
P. Sorba$^{b}$\footnote{sorba@lapp.in2p3.fr} and 
Ph. Zaugg$^{c}$\footnote{zaugg@CRM.UMontreal.CA}}\\

\null

\noindent
{\it $^a$ INFN, Dipart. di Fisica dell'Univ. di
  Pisa, Piazza Torricelli 2, 56100 Pisa, Italy\\[2.1ex]
$^b$ LAPTH, Chemin de Bellevue, BP 110, F-74941 Annecy-le-Vieux
  cedex, France\\[2.1ex]
$^c$ CRM, Univ. de Montr{\'e}al, CP 6128 
Succ. Centre-ville, Montr{\'e}al 
(Qc) H3C 3J7, Canada}

\vfill

\begin{abstract}
We study the Yangian symmetry of the multicomponent Quantum Nonlinear Schr\"odinger
hierarchy in the framework of the Quantum Inverse Scattering Method. We
give an explicit realization of the Yangian generators in
terms of the deformed oscillators algebra which naturally occurs in
this framework.
\end{abstract}

\vfill
\rightline{LAPTH-733/99}
\rightline{IFUP-TH 24/99}
\rightline{CRM-2608}
\rightline{\tt hep-th/9905105}
\newpage
\pagestyle{plain}
\setcounter{page}{1}
\sect{Introduction}

\null

An increasing number of integrable systems with internal degrees of
freedom have been shown to exhibit a Yangian symmetry.  One of the
earliest example of this is perhaps the Haldane-Shastry spin
chain~\cite{BGHP} and the spin generalisation of the
Calogero-Sutherland model investigated in~\cite{ABB}. 

In the realm of integrable systems the quantum nonlinear
Schr{\"o}dinger (NLS) model distinguishes itself by being one of the
most studied system and its simplest version played an important role
in the development of the Quantum Inverse Scattering Method (QISM).
Recently the authors of~\cite{Wad} have considered the quantum NLS
model with spin $\frac{1}{2}$ fermions and repulsive interaction on
the line and have unraveled the presence of a Yangian symmetry
$Y(sl(2))$. 

In this paper we consider the most general case of the quantum NLS
model with $N$-components bosons or fermions and prove that the
Yangian symmetry is \Ysl.  In addition we provide an explicit
realisation of the Yangian generators using the algebra of creation
and annihilation operators for scattering states that is an essential
part of the QISM, also known as the Zamolodchikov-Faddeev (ZF)
algebra~\cite{ZF}.  This approach makes it clear that the Yangian is
actually a symmetry of the whole quantum integrable hierarchy whose
lowest instance is the NLS model. As a by-product, we also obtain all
the solutions to the equations of motion of the quantum NLS hierarchy.

The structure of the paper is the following.  In section 2 we
summarize the QISM applied to the NLS model.  Section 3 in devoted to
the study of the Yangian symmetry of the NLS model.  In section 4 we
outline the construction of the Yangian generators, as well as the
higher Hamiltonians of the NLS hierarchy, in terms of creation and
annihilation operators of the ZF algebra, the technical details being
gathered in the appendices.  Next we present a connection between the
action of the Yangian algebra restricted to the subspaces of fixed
particle number 
and a class of finite $W$-algebras.  We conclude with possible extensions of this work.

\sect{The multicomponent nonlinear Schr\"odinger model}

We shall consider the $N$-components quantum nonlinear Schr\"odinger model with bosons or fermions and repulsive coupling.  We first collect a few known results from the Quantum Inverse Scattering Method applied to the model, most of which can be found in \cite{PWZ,Gut} with more details.

The Hamiltonian of the NLS model is
\be
H = \int dx \left(\frac{\partial \phi^{\dag i}}{\partial x} \frac{\partial \phi_i}{\partial x} - \rho g \phi^{\dag i} \phi^{\dag j} \phi_i \phi_j \right),
\label{hamiltonian}
\ee
where the fields operators satisfy the equal time canonical commutation relations
$$
[ \phi_i(x), \phi^{\dag j}(y)]_\rho = \delta_i^j \delta(x-y),
\qquad
[ \phi_i(x), \phi_j(y)]_\rho = [ \phi^{\dag i}(x), \phi^{\dag j}(y)]_\rho = 0.
$$
The conventions are that $\rho=-1$ for bosons, $\rho=1$ for fermions
and $[\;,\;]_\rho$ stands respectively for the commutator or anti-commutator.  Latin indices run from 1 to $N$, whereas Greek indices run from 1 to $N+1$.  Repeated indices are summed over their appropriate range.

The linear operator of the QISM is
\be
L(x|\lambda) = i \frac{\lambda}{2} \Sigma + \Omega(x), \quad\hbox{with}\quad \Omega(x)= i \sqrt{g} \left(\phi_j(x) E_{j,N+1} - \phi^{\dag j}(x) E_{N+1,j}\right).
\ee
In this equation $E_{\alpha \beta}$ is the standard $(N+1) \times (N+1)$ matrix $(E_{\alpha \beta})_{\mu\nu} = \delta_{\alpha \mu} \delta_{\beta \nu}$ and $\Sigma$ is a diagonal matrix $\Sigma = I_{N+1}-2 E_{N+1,N+1}$, where $I_{N+1}$ is the identity matrix.

The quantum monodromy matrix $T(x,y|\lambda)$ is defined by the equations
\be
\frac{\partial}{\partial x} T(x,y|\lambda) = \ :\! L(x|\lambda) T(x,y|\lambda)\! :,
\qquad
T(x,y|\lambda)\vert_{x=y}=I_{N+1},
\label{dTLT}
\ee
where $:~:$ denotes the usual normal order for the field operators
$\phi(x)$ and $\phi^\dag(x)$.  The infinite volume limit is a delicate
issue in the QISM~\cite{Gut}. The infinite volume monodromy matrix
$T(\lambda)$ is formally defined by
$$
T(\lambda) = \lim_{x \to \infty,y \to -\infty} E(-x|\lambda) T(x,y|\lambda) E(y|\lambda),
$$
where $E(x|\lambda)=\exp(i \lambda x \Sigma/2)$.
Using the implicit representation for $T(x,y|\lambda)$,
$$
T(x,y|\lambda) = E(x-y|\lambda) + \int_y^x dx_1 \,  E(x-x_1|\lambda) : \! \Omega(x_1) T(x_1,y|\lambda)\! :
$$
the monodromy matrix $T(\lambda)$ can formally be computed through an iterative procedure and is expressed as
\be
T(\lambda) = I_{N+1} + \sum_{n=1}^\infty \int_{-\infty}^\infty d^nx \ \theta(x_1 > \dots > x_n) E(2 \sum_{i=1}^n (-)^i x_i|\lambda) :\!\Omega(x_1) \dots \Omega(x_n)\! :
\label{iteT}
\ee

The commutation relations for infinite volume are encoded in the exchange relation
\be
R_\rho^+(\lambda-\mu)\ T(\lambda) \rhotimes T(\mu) =
T(\mu) \rhotimes T(\lambda)\
R_\rho^-(\lambda-\mu)
\label{RTT}
\ee
with the $R$-matrices
\bea
 R_\rho^\pm(\mu) &=& \frac{i\rho g}{\mu+i\rho g}\pv\frac{1}{\mu} \ E_{jj}\otimes E_{kk}+ \frac{1}{\mu+i\rho g}\ E_{\alpha j}\otimes E_{j\alpha} \nonumber\\
 &+&\frac{\mu-i\rho g}{(\mu+i0)^2}\ E_{j,N+1}\otimes E_{N+1,j} -\frac{\rho(\mu-i g)}{\mu+i\rho g}\pv\frac{1}{\mu}\ E_{N+1,N+1}\otimes E_{N+1,N+1}\nonumber\\
 &\pm& i\pi\delta(\mu)(E_{jj}\otimes E_{N+1,N+1}
-E_{N+1,N+1}\otimes E_{jj}).
\eea
The term $i0$ is a consequence of the principal value regularisation adopted when $\mu$ goes to zero, according to the relation (in the sense of distributions)
$$
\pv\frac{1}{\mu} = \frac{1}{\mu \pm i0} \pm i \pi \delta(\mu).
$$

It will be convenient to rename some elements of the monodromy matrix such as $D(\lambda)=T_{N+1,N+1}(\lambda)$ and $b^j(\lambda)=T_{N+1,j}(\lambda)$.  Further examination of some components of (\ref{RTT}) yields the following relations
\bea
D(\lambda) D(\mu) &=&  D(\mu) D(\lambda), \label{DD} \\
D(\lambda) b^j(\mu) &=& \frac{\lambda-\mu+ig}{\lambda-\mu+i0}\ b^j(\mu) D(\lambda), \label{Db}\\
b^j(\lambda) b^k(\mu)& =&
\frac{-\rho(\lambda-\mu)}{\lambda-\mu-ig}\ b^k(\mu)b^j(\lambda)-
\frac{ig}{\lambda-\mu-ig}\ b^j(\mu)b^k(\lambda).
\label{bb}
\eea
The matrix element $D(\lambda)$ serves as a generating operator-function for the commuting integrals of motion of the NLS model.  This is most easily seen by performing an asymptotic expansion for large $\lambda$ in the solution (\ref{iteT}), whereby one gets
\bea
 D(\lambda) &=& 1 + \frac{i g}{\lambda} \hat{N} + \frac{i g}{\lambda^2} \left( P+ i \frac{g}{2} \hat{N}(\hat{N}-1)\right) \label{Dexp} \\
 &+& \frac{i g}{\lambda^3} \left(H + i g (\hat{N}-1)P - \frac{g^2}{6}\hat{N}(\hat{N}-1)(\hat{N}-2) \right) + O(\frac{1}{\lambda^4})
\nonumber
\eea
with
$$
\hat{N} = \int dx \ \phi^{\dag j} \phi_j, \qquad
P = - i \int dx \ \phi^{\dag j} \partial\phi_j ,
$$
and $H$ is given in (\ref{hamiltonian}).
Consequently, eq.~(\ref{DD}) implies that these integrals of motion are all in involution.

In the multicomponent NLS model the commutation relations amongst $b^j(\lambda)$ and their adjoint have to be deduced from another type of exchange relation (see \cite{PWZ} for details).  Moreover these operators hardly make sense as operators on the Hilbert space~\cite{Gut} and it is necessary to consider instead the scattering states operators
\be
a^{\dag j}(\lambda) = \frac{i}{\sqrt{g}} b^j(\lambda) D^{-1}(\lambda),
\label{defa}
\ee
and their adjoint $a_k(\lambda)$.
Then the commutation relations amongst $a(\lambda)$ and $a^\dag(\lambda)$ are nicely encoded in the form of a Zamolodchikov-Faddeev (ZF) algebra \cite{ZF} as (a sort of deformed oscillator algebra):
\bea
a_j(\lambda) a_k(\mu) &=& R_{kj}^{nm}(\mu-\lambda)\ a_m(\mu) a_n(\lambda), \\
a^{\dag j}(\lambda) a^{\dag k}(\mu) &=&  a^{\dag m}(\mu) a^{\dag n}(\lambda)\ R_{mn}^{jk}(\mu-\lambda),\\
a_j(\lambda) a^{\dag k}(\mu) &=&  a^{\dag m}(\mu)\ R_{jm}^{kn}(\lambda-\mu)\ a_n(\lambda) + \delta_j^k \delta(\lambda-\mu),
\eea
where the $R$-matrix given by
\be
R_{jm}^{kn}(\mu) = \frac{1}{\mu+ig} \left( -\rho \mu\ \delta_m^k \delta_j^n + ig\ \delta_j^k \delta_m^n \right)
\label{RmatZF}
\ee
is the two-body scattering matrix of the $N$-components NLS model.

The operators $a^{\dag}(\lambda)$ and $a(\lambda)$ play the role of creation and annihilation operators and as such can be used to build a Fock space through their action on the vacuum.  Owing to (\ref{Db}) these states are simultaneous eigenstates of the conserved quantities.  Actually, the set of (asymptotic scattering) states $\prod^n_{i=1} a^{\dag i}(\lambda_i)|0 \rangle$ for all $n$ is dense in the Hilbert space of the NLS model~\cite{Gut,fock}.  This property will be useful later on.

The original field operator $\phi_k(x)$ can be recovered from the
knowledge of the scattering data.  This is achieved by solving a
system of quantum Gel'fand--Levitan equations, whose result is a
quantum version of the Rosales expression~\cite{Ros,solQ} (given here for $t=0$):
\pagebreak[3]
\bea
\phi_k(x) &=& \sum_{n=0}^\infty (-g)^n \int \frac{d^n\nu\ d^{n+1}\mu}{(2\pi)^{2n+1}} a^{\dag i_1}(\nu_1) \dots a^{\dag i_n}(\nu_n) a_{i_n}(\mu_n) \dots a_{i_1}(\mu_1) a_k(\mu_0) \nonumber \\
&&\frac{e^{i \mu_0 x} \prod_{l=1}^n e^{i(\mu_l-\nu_l) x}}{\prod_{l=1}^n (\nu_l - \mu_{l-1}+i\epsilon)(\nu_l - \mu_{l}+i\epsilon)}.
\label{solutionphi}
\eea

\sect{Yangian algebra in NLS}

In this subsection we shall show that the NLS model contains an infinite set of conserved charges having the structure of a Yangian algebra.  There are several equivalent definitions of the Yangian \Ysl\ and we present two of them in Appendix~A.

The Yangian symmetry of the NLS model manifests itself already in the exchange relation (\ref{RTT}).  Indeed, denoting by $\Tt(\lambda)$ the $N \times N$ submatrix of $T(\lambda)$, $\Tt(\lambda)=T_{ij}(\lambda) E_{ij}$, and examining the appropriate components of (\ref{RTT}), one deduces the following relations
\be
\widetilde{R}(\lambda-\mu) \ \Tt(\lambda) \otimes \Tt(\mu) =
 \Tt(\mu) \otimes \Tt(\lambda) \ \widetilde{R}(\lambda-\mu)
\label{RTtTt}
\ee
with yet another $R$-matrix
\be
\widetilde{R}(\lambda-\mu) = 
(\lambda-\mu) E_{jk} \otimes E_{kj} +i \rho g I_N \otimes I_N  .
\label{Rtilde}
\ee
This coincides precisely with the defining relation of the Yangian \Ygl.

The fact that the Yangian algebra commutes with the Hamiltonian of the NLS model is a consequence of the exchange relation as well, since one extracts from (\ref{RTT}) that
\be
[\Tt_{ij}(\lambda),D(\mu)] = 0
\label{TDzero}
\ee
and the Hamiltonian is just one of the conserved quantities in the asymptotic expansion (\ref{Dexp}).

It is of some interest to get an explicit representation of the Yangian generators in terms of the field operators $\phi_i(x)$.  This is achieved with the help of the iterative solution (\ref{iteT}) of the monodromy matrix $T(\lambda)$, and by looking at its asymptotic expansion for large $\lambda$ one finds that
\bea
 \Tt_{jk}(\lambda) &=& \delta_{jk} + \frac{i \rho g}{\lambda} \int dx \ \phi^{\dag k}(x) \phi_j(x) + \frac{\rho g}{\lambda^2} \Big( \int dx\ \phi^{\dag k}(x) \partial\phi_j(x) \label{expTij} \\
 && \mbox{}+ g\int d^2x \theta(x_1>x_2) \phi^{\dag n}(x_1) \phi^{\dag k}(x_2) \phi_j(x_1) \phi_n(x_2) \Big) + O(\frac{1}{\lambda^3})\nonumber \\
 & \equiv & \delta_{jk} + i \rho g \sum_{n=0}^\infty \frac{\Tt_{jk}^{(n)}}{\lambda^{n+1}} .
\eea
In particular this shows that in this model the formal series expansion of the Yangian generators (see \ref{formalt}) is to be understood as an asymptotic expansion for large $\lambda$.

The most important relations for our purpose are the commutators of the Yangian generators $\Tt_{jl}^{(n)}$ with the creation operators $a^{\dag k}(\mu)$.  Using the exchange relation (\ref{RTT}), the definition (\ref{defa}) and the symmetry property (\ref{TDzero}) one finds that
\be
[ \Tt_{jl}(\lambda), a^{\dag k}(\mu) ] = \frac{i \rho g}{\lambda-\mu-i0} a^{\dag l}(\mu) \Tt_{jk}(\lambda),
\label{Ta}
\ee
which, upon expanding in $1/\lambda$, yields
\bea
 {[\Tt_{jl}^{(0)},a^{\dag k}(\mu)]} &=& a^{\dag l}(\mu) \delta_{jk}, \label{Tna*}\\
 {[\Tt_{jl}^{(1)},a^{\dag k}(\mu)]} &=& \mu \,  a^{\dag l}(\mu) \delta_{jk} + i \rho g \, a^{\dag l}(\mu) \Tt_{jk}^{(0)} .\nonumber
\eea

In the next section we shall need a more convenient basis of \Ysl\ (see Appendix~A).  It will also be convenient to deal with self-adjoint generators for the Yangian algebra.  It turns out that these two requirements can be fulfilled in a single operation.  Let us denote by $\Tt(\lambda)^\ddag$ the hermitian conjugate of the $N \times N$ matrix $\Tt(\lambda)$ obtained by transposing the matrix and taking the adjoint of its entries (which are operators in a Hilbert space), namely
\be
\Tt(\lambda)^\ddag = I_N - i \rho g \sum_{n=0}^\infty \frac{\Tt^{(n) \dag}_{jk} E_{kj}}{\lambda^{n+1}}
.
\ee
Its anti-hermitian part is simply
\be
\frac{1}{2}(T(\lambda)-T(\lambda)^\ddag) = i \rho g \sum_{n=0}^\infty  \frac{1}{\lambda^{n+1}} \frac{1}{2} \left(\Tt^{(n)}_{jk} + \Tt^{(n) \dag}_{kj} \right)E_{jk} \equiv i \rho g \, U(\lambda),
\ee
where $U(\lambda)=U(\lambda)^\ddag$ is now hermitian.  As such it can be expanded as
\be
U(\lambda) = \sum_{n=0}^\infty \frac{U^{(n)}_{jk} E_{jk}}{\lambda^{n+1}} = \sum_{n=0}^\infty \frac{\sum_{a=1}^{N^2-1}\Qt_{n}^a t_a + \Qt_{n}^0 I_N}{\lambda^{n+1}},
\ee
where $\Qt_{n}^a = \Qt_{n}^{a\dag}$ are self-adjoint generators.  The matrices $t_a = (t_a)^{*T}$ are in the fundamental representation of $su(N)$ and normalized to
$$
[t_a,t_b] = i {f_{ab}}^c t_c, 
\qquad
\eta_{ab} = tr( t_a t_b).
$$
Therefore we have that
\be
\Qt_{n}^a = tr( U^{(n)} t^a) = U^{(n)}_{jk} (t^a)_{kj}.
\label{defQtn}
\ee
In the $\Qt^a_n$ basis, only the $n=0,1$ grades are necessary and from (\ref{expTij}) we find that
\bea
 U^{(0)}_{jk} &=&\Tt^{(0)}_{jk}, \label{U0U1} \\
 U^{(1)}_{jk} &=& \Tt^{(1)}_{jk} - \frac{i \rho g}{2}\, \Tt^{(0)}_{lk}\Tt^{(0)}_{jl}  + \frac{i \rho g N}{2}\, \Tt^{(0)}_{jk}. \nonumber
\eea
The commutation relations of $\Qt_{n}^a$ and $a^{\dag k}(\mu)$ are then readily computed using (\ref{Tna*}), (\ref{defQtn}) and (\ref{U0U1})
\bea
 {[\Qt_0^a,a^{\dag k}(\mu)]} &=& (a^{\dag}(\mu) t^a)^k, \label{Q0a}\\
 {[\Qt_1^a,a^{\dag k}(\mu)]} &=& \mu\, (a^{\dag}(\mu) t^a)^k-\frac{\rho g}{2}
{f^a}_{bc} (a^{\dag}(\mu) t^c)^k \Qt^b_{0}. \label{Q1a}
\eea

The operators $\Qt^a_{0,1}$ generates the Yangian \Ysl\ and the operators $\Qt^0_n$ are related to its center.  They are also connected in some intricate way to the integrals of motion.

It is instructive to study the type of Yangian representations that appear in the Hilbert space of the NLS model.  The vacuum is invariant under the action of $\Qt_{0,1}^a$ and the one-particle state $a^{\dag k}(\mu) |0 \rangle$ transforms as an evaluation representation in the fundamental representation of $sl(N)$~\cite{CP}, denoted by $V(\mu)$, since
\beano
 \Qt_0^a \ a^{\dag k}(\mu) |0 \rangle &=& {(t^a)_j}^k a^{\dag j}(\mu)|0 \rangle,\\
 \Qt_1^a \ a^{\dag k}(\mu) |0 \rangle &=& \mu\, {(t^a)_j}^k a^{\dag j}(\mu)|0 \rangle .
\eeano
The two-particles state $a^{\dag k_1}(\mu_1) a^{\dag k_2}(\mu_2) |0 \rangle$ transforms as a tensor product of two such representations
\bea
 \Qt_0^a \ a^{\dag k_1}(\mu_1) a^{\dag k_2}(\mu_2) |0 \rangle &=& ({(t^a)_{j_1}}^{k_1} \delta_{j_2}^{k_2} + \delta_{j_1}^{k_1} {(t^a)_{j_2}}^{k_2})a^{\dag j_1}(\mu_1) a^{\dag j_2}(\mu_2) |0 \rangle , \nonumber\\
 \Qt_1^a \ a^{\dag k_1}(\mu_1) a^{\dag k_2}(\mu_2) |0 \rangle &=& \big(\mu_1 \,{(t^a)_{j_1}}^{k_1} \delta_{j_2}^{k_2} + \mu_2 \,\delta_{j_1}^{k_1} {(t^a)_{j_2}}^{k_2} \label{repVV}\\
 &&\mbox{} -\frac{\rho g}{2} {f^a}_{bc} {(t^c)_{j_1}}^{k_1} {(t^b)_{j_2}}^{k_2}\big)  a^{\dag j_1}(\mu_1) a^{\dag j_2}(\mu_2) |0 \rangle . \nonumber
\eea
In particular, the second term on the rhs of (\ref{Q1a}) ensures that the action of $\Qt_{1}^a$ on the tensor product is consistent with the comultiplication of the Yangian
\beano
 \Delta(\Qt_0^a) &=& \Qt_0^a \otimes 1 + 1 \otimes \Qt_0^a ,\\
 \Delta(\Qt_1^a) &=& \Qt_1^a \otimes 1 + 1 \otimes \Qt_1^a - \frac{\rho g}{2} {f^a}_{bc} \Qt_0^c \otimes \Qt_0^b .
\eeano
Therefore $n$-particles states will carry a $n$-fold tensor product of $V(\mu_i)$ representations.

\sect{Yangian generators and deformed oscillators}

In view of formul{\ae} (\ref{solutionphi}) and (\ref{expTij}), it is
clear that trying to reconstruct the Yangian generators (in term of
oscillators) by direct calculation is a difficult task. Instead we
shall define two operators $Q_0^a$ and $Q_1^a$ that have the same
commutation relations with $a^\dag(\mu)$ as in (\ref{Q0a}) and
(\ref{Q1a}).  Therefore their action on the Fock space spanned by the
$a^\dag(\mu)$ will coincide with that of $\Qt_{0,1}^a$ and, as this
Fock space is dense in the Hilbert space of the NLS model, we shall
identify these operators.  All the other Yangian generators are built
from these two sets of elements.

In order to simplify the presentation, we adopt a more compact notation for the ZF algebra.  We drop the explicit mention of the indices $i,j$ and the momenta $\mu_i$, and introduce instead a new index refering to an $N$-dimensional auxiliary space.  More explicitly
$$
a_1 \equiv a_i(\mu_1) \, e_1^i,
$$
where $e_1^i$ is some basis of the $N$-dimensional auxiliary space labeled by 1.
For instance, the $R$-matrix (\ref{RmatZF}) reads
\be
R_{12} \equiv R_{12}(\mu_1-\mu_2) = \frac{1}{\mu_1-\mu_2+ ig}(-\rho (\mu_1-\mu_2) 1 \otimes 1 +ig P_{12}),
\ee
where $P_{12}$ is the permutation operator in the auxiliary spaces.  The inverse of $R_{12}$ is $R_{21}=R_{21}(\mu_2-\mu_1)$.  With this notation, the ZF algebra relations read
\bea
a_1 a_2 &=& R_{21} \, a_2 a_1, \nonumber \\
a_1^\dag a_2^\dag &=&  a_2^\dag a_1^\dag \, R_{21},  \label{ZFa}\\
a_1 a_2^\dag &=& a_2^\dag \, R_{12} \, a_1 + \delta_{12} .\nonumber
\eea
We also rename the operators we are looking for as $J^a=Q_0^a$ and $S^a=Q_1^a$.  Thus (\ref{Q0a}) and (\ref{Q1a}), and their conjugate, translate to
\bea
{[J^a, a^\dag_0]} &=& a^\dag_0 t^a_0, \label{Ja*}\\
{[J^a, a_0]} &=& - t^a_0 a_0, \label{Ja}\\
{[S^a, a^\dag_0]} &=&\mu_0\, a^\dag_0 t^a_0 + \frac{\rho g}{2} {f^{a}}_{bc} a^\dag_0 t^b_0 J^c, \label{Sa*} \\
{[S^a, a_0]} &=& -\mu_0\, t^a_0 a_0 + \frac{\rho g}{2} {f^{a}}_{bc} J^b t^c_0 a_0. \label{Sa}
\eea
Here $t_0^a$ means that the $su(N)$ matrix $t^a$ is acting in the auxiliary space labeled by $0$.

We first construct the operator $J^a$. The idea is to built it
recursively in such a way that the relations (\ref{Ja*}-\ref{Ja})
are fulfilled. The expansion parameter is not the coupling
constant $g$ but rather the number of oscillators. More precisely, we start
with the expression
\be 
J^a = \sum_{n=1}^\infty \frac{(-)^{n+1}}{n!} J^a_{(n)} 
= \sum_{n=1}^\infty \frac{(-)^{n+1}}{n!} a^\dag_{1\dots n} 
T^a_{1\dots n} a_{n\dots 1},
\label{defJ}
\ee
where $a_{n\dots 1}=a_n(\mu_n) \cdots a_2(\mu_2) a_1(\mu_1)$ and the integration on $\mu_1, \mu_2,\dots,\mu_n$ is implied in $J^a$. 
We then determine the tensors $T^a_{1\dots n}$ recursively. 
The details of the calculation are relegated in Appendix~B, and we find as a result that
\be
T^a_{1\dots n}= \sum_{j=1}^n \alpha_j^n\ t^a_j
\qquad \mbox{with}
\qquad \alpha_j^n=
(-)^{j-1}
\bin{n-1}{j-1}.
\label{defT}
\ee
These generators verify as well
\be
{[J^a, J^b]} = i{f^{ab}}_c J^c.
\ee
They form the $sl(N)$ subalgebra of \Ysl.
\null

We then look for operators $S^a$ of a similar form
\be
S^a = \sum_{n=1}^\infty \frac{(-)^{n+1}}{n!} S^a_{(n)} 
= \sum_{n=1}^\infty \frac{(-)^{n+1}}{n!} a^\dag_{1\dots n} 
\wT^a_{1\dots n} a_{n\dots 1}
\label{defS}
\ee
satisfying (\ref{Sa*}-\ref{Sa}).  In this case, the procedure is simpler since $S^a$ lives in the adjoint representation of the subalgebra generated by $J^a$.  This implies that
\be
S^a = -\frac{i}{c_2} {f^a}_{bc} \, [S^b, J^c],
\label{SfSJ}
\ee
where $c_2$ is as usual the second Casimir in the adjoint
representation, $c_2 \delta_a^{b}= f_{acd} f^{bcd}$.  Using the explicit expression (\ref{defJ}) for $J^a$ and imposing that $S^a$ satisfies (\ref{Sa*}-\ref{Sa}) enables us to compute the right-hand side of (\ref{SfSJ}) and to determine the tensors $\wT^a_{1\dots n}$.

As anticipated from (\ref{Sa*}) the tensors depend on the momenta and their expression is (see Appendix C):
\be
\wT^a_{1\dots n}=\sum_{j=1}^n \alpha^n_j\,\left(
\mu_j\, t^a_j\, +\frac{\rho g}{2}{f^{a}}_{bc}\, 
\sum_{i=1}^{j-1}  t^{b}_i t^c_j\right) .
\ee
Therefore the expressions (\ref{defJ}) and (\ref{defS}) provide a realisation of the Yangian generators in terms of the ZF algebra.

A similar procedure can be applied to the integrals of motion (or higher Hamiltonians) of the NLS model.  The lowest Hamiltonians $\hat{N}, P, H$ are explicitly known, and from (\ref{Db}), (\ref{Dexp}) and (\ref{defa}) we find that
\beano
{[\hat{N},a^\dag_0]} &=& a^\dag_0 ,\\
{[P,a^\dag_0]} &=& \mu_0 \ a^\dag_0 ,\\
{[H,a^\dag_0]} &=& \mu_0^2 \ a^\dag_0 .
\eeano
Let us now define the infinite set of commuting operators $\hI_n, n \geq 0$ by
\be
\hI_n = \int\! d\mu_1 \ \mu_1^n a_1^{\dag} a_1
\ee
which enjoy the following commutation relations
\bea
{[\hI_n,a^\dag_0]} &=& \mu_0^n \ a^\dag_0 ,\label{Ina*a} \\
{[\hI_n,a_0]} &=& -\mu_0^n \ a_0 .\nonumber
\eea
This implies that acting on a $m$-particles state
\be
\hI_n \, a_1^\dag \dots a_m^\dag |0 \rangle = (\sum_{j=1}^m \mu_j^n) \, a_1^\dag \dots a_m^\dag |0 \rangle
\ee
which is precisely the definition of the higher Hamiltonians in the quantum NLS model~\cite{Gut}.  According to (\ref{Ina*a}) the lowest ones are obviously identified with
$$
\hat{N} = \hI_0,
\qquad
P = \hI_1,
\qquad
H = \hI_2.
$$

Moreover we can show that $D(\lambda)$ is a generating
operator-function for the integrals of motion, that is, it can be
expressed entirely in terms of the $\hI_n$ operators.  Indeed we can prove that
\be
D(\lambda) = \exp(d(\lambda)),
\qquad
\hbox{where}
\qquad
d(\lambda) = \sum_{n=0}^\infty \ \frac{d_n}{\lambda^{n+1}},
\label{Dexpd}
\ee
and
\be
d_n = ig \sum_{j=0}^n \ \frac{(-i g)^{n-j}}{n+1} 
\bin{n+1}{j} \ \hI_j.
\label{dn}
\ee
As the $\hI_n$ commute with each other and satisfy (\ref{Ina*a}), it is straightforward to show that 
$$
\exp(d(\lambda)) \ a_0^\dag \ \exp(-d(\lambda)) = \left( 1 + \frac{i g}{\lambda-\mu_0)} \right) a_0^\dag
$$
which is precisely the relation between $D(\lambda)$ and $a_0^\dag$ as deduced from (\ref{Db}).  This proves the assertion (\ref{Dexpd}).

Owing to the explicit expressions (\ref{defJ}) and (\ref{defS}), it is
very easy to check that the operators $\hI_n$ commute with $J^a$ and
$S^a$.  As $J^a, S^a$ generate the Yangian algebra, then obviously the
$\hI_n$ commute with the whole Yangian algebra.  This is just another
way of expressing the content of (\ref{TDzero}). It also means that
the Yangian is a symmetry of all the quantum systems defined with the
help of the higher Hamiltonians.

As usual, the time evolution of the quantum field $\phi_k(x)$
of eq. (\ref{solutionphi}) is given by the conjugation by the NLS
Hamiltonian (\ref{hamiltonian}) which is nothing but
$\hI_2$. According to the commutation relations (\ref{Ina*a}), this
amounts to multiply in (\ref{solutionphi}) the creation operators
$a^\dag(\nu)$ by $e^{i\nu^2 t}$ and the annihilation operators
$a(\mu)$ by $e^{-i\mu^2 t}$. Although the expression
(\ref{solutionphi}) was originally obtained as the solution to the NLS
equation of motion, it also provides the solution to the higher flows
of the hierarchy. Simply, the $t_n$-time evolution is now induced by a
conjugation by $\exp(i\hI_n t_n)$. Consequently, the phases
multiplying the creation and annihilation operators are respectively $e^{i\nu^n t_n}$
and $e^{-i\mu^n t_n}$. It is remarkable to get in such an easy way the
solutions to all the equations of motion of the hierarchy.

\sect{Connection with finite $W$-algebras}

It has already been shown that there is a strong connection between
Yangians \Ysl\ and finite $W$ algebras~\cite{13}. We show in this section
that the NLS hierarchy offers a natural framework to illustrate this
relation.

For such a purpose, we focus on the Fock space $\cal F$
spanned by the $a^{i \dag}(\mu)$ and detail its structure.  Let us
recall that the $p$-particles subspace ${\cal F}_p(\mu_1,\dots,\mu_p)$
with fixed momenta $(\mu_1,\dots,\mu_p)$ is a tensor product of $p$
evaluation representations $\otimes_{i=1}^p V(\mu_i)$, all in the
fundamental representation of $su(N)$.  The $p$-particles subspace
${\cal F}_p$ is just the span over all momenta $\mu_i$.

On ${\cal F}_p$ it is straightforward to see that the Yangian
generators $t_{ij}^{(n)}$ with $n\geq p$ act as zero operators.  We thus
have a representation of a truncated Yangian, which is known to be
isomorphic, at the algebra level, to a finite $W(gl(Np),N.sl(p))$
algebra~\cite{BR}.  Thus on each $p$-particles subspace the Yangian
acts as a $W(gl(Np),N.sl(p))$ algebra.  This is another nice application of
finite $W$ algebras (more examples can be found in~\cite{BRS} and
references therein).

Let us illustrate this point on the simplest case, namely $p=2$ and
$N=2$.  Consequently, the only independent Yangian generators are
$Q_0^a$, $Q_1^a$ and their action on ${\cal F}_2$ is given in
(\ref{repVV}).  Let us denote their representation on ${\cal F}_2$ by
$J^a$ and $S^a$ respectively.  Up to an innocuous shift $S^a \to S^a -
\frac{1}{2}P J^a$, the full set of commutation relations satisfied by
these operators is
\bea
{[J^a,J^b]} & = & i {f^{ab}}_c \ J^c, \nonumber\\
{[J^a,S^b]} & = & i {f^{ab}}_c \ S^c, \\
{[S^a,S^b]} & = & i {f^{ab}}_c \ \left( \frac{1}{2}H - \frac{1}{4} P^2
  - g^2 \frac{c_2}{8N} \right) J^c. \nonumber
\eea
One recognizes here the relations of the $W$-algebra
$W(gl(4),2.sl(2))$. 

$W(gl(4),2.sl(2))$ and $W(sl(4),2.sl(2))$ algebras
essentially differ by one central element. In the present context,
this element is nothing but $P$, the total momentum.
Thus, the transition between the two $W$-algebras amounts to describe
the system in its center of mass frame. 

\sect{Conclusions and outlook}

We deliberately imposed the restriction that the coupling constant $g$
be positive.  When $g$ is negative, the quantum spectrum also contains
bound states (solitons) and the asymptotic scattering states
$\prod^n_{i=1} a^{\dag i}(\lambda_i)|0 \rangle$ are no longer complete
in the Hilbert space of the NLS model.  Our construction relies
crucially on the completeness of those states in order to identify the
generators of the Yangian symmetry (\ref{defQtn}) with the generators
(\ref{defJ}) and (\ref{defS}) expressed in term of the Zamolodchikov-Faddeev algebra elements.

In both regions of the coupling constant, the Yangian generators
commute with the scattering matrix and the Yangian corresponds to a
symmetry of the scattering matrix.  The nuance is that in the second
situation we have only defined the action of the Yangian generators on
the asymptotic scattering states and the best we can say about the
operators (\ref{defJ}) and (\ref{defS}) is that they generate an
asymptotic symmetry of the NLS model.  Extending their definition to
the full Hilbert space would require the complete knowledge of the
bound states spectrum of the NLS model, something not yet achieved. 

From a more general point of view, when considering our construction
as solely based on the existence of a ZF algebra we may conclude that
it is possible to realise a Yangian algebra on the Fock space
generated by the ZF algebra whenever the $R$-matrix is of the rational type given in~(\ref{RmatZF}).  For the NLS model, the $S$-matrix is invariant under $SU(N)$ and the symmetry algebra turns out to be a Yangian \Ysl.  Let us mention that the $SU(N)$-Thirring model is another quantum system with such rational $R$-matrix, thus providing a relativistic example of a system with Yangian symmetry~\cite{Aris}.

It would be interesting to exhibit the symmetry algebra of more
general cases where the $R$-matrix is not rational but is still
invariant under some Lie group.  As the $R$-matrix in the ZF algebra
can be interpreted as the two-body $S$-matrix of integrable systems in
1+1 dimensions, this question reduces to the problem of identifying
the largest $S$-matrix symmetry of such integrable systems.  The study
of this sort of relationship in a more general situation is certainly
interesting and we intend to return to it in a future paper. 

The NLS model considered in the present work is defined on the line.
Usually the presence of boundaries strongly influences the symmetry
and we are currently investigating this issue in the NLS model on the
half line.  Its quantification as a quantum field theory, as carried
out in~\cite{GLM}, reveals the presence of a boundary exchange algebra
generalizing the ZF algebra, thus fitting well within the more general
scheme of integrable systems with boundaries developed
in~\cite{Bound}.  We still expect the system to possess a large
internal symmetry, possibly in the form of a twisted Yangian
algebra~\cite{Molev} and we also expect to be able to express these
symmetry generators in terms of the boundary exchange
algebra~\cite{MRSZ2}.

\null

\noindent{\bf Acknowledgements:}
P.S. would like to express his gratitude to the members of CRM in Montreal for
their kind invitation for an extended visit this spring. Ph.Z. would
like to thank L. Vinet and Y. Saint-Aubin for interesting discussions.

\appendix
\sect{Definition of \Ygl}
The Yangian \Ygl\ can be defined as the free associative algebra over
$\mathbf C$ with generators $1, t_{ij}^{(n)}, n \geq 0$ (not to be confused
with $t^a_k$ used elsewhere in the text) quotiented by the relation~\cite{CP}
\be
R_Y(\lambda-\mu) \ t(\lambda) \otimes t(\mu) = 
t(\mu) \otimes t(\lambda) \ R_Y(\lambda-\mu),
\ee
where we introduced the $N \times N$ matrix $t(\lambda)$ whose entries are formal series in $\lambda$
\be
t_{ij}(\lambda) = \delta_{ij} + h \sum_{n=0}^\infty \frac{t_{ij}^{(n)}}{\lambda^{n+1}}
\label{formalt}
\ee
and the $R$-matrix is given by
\be
R_Y(\lambda-\mu) = (\lambda-\mu) E_{ij} \otimes E_{ji} - h I_N \otimes I_N .
\ee
The non-zero deformation parameter $h$ can be scaled away as two Yangians with different non-zero deformation parameter are known to be isomorphic.

The quantum determinant
\be
det_q(t(\lambda)) = \sum_{\pi \in S_N} sign(\pi) t_{1,\pi(1)}(\lambda-N+1) \dots t_{N,\pi(N)}(\lambda)
\ee
generates the infinite dimensional center $\cZ$, and the Yangian \Ygl\ is isomorphic to $\cZ \otimes Y(sl(N))$.

The coproduct in this presentation is simply
\be
\Delta( t_{ij}(\lambda) ) = \sum_{k=1}^N t_{ik}(\lambda) \otimes
t_{kj}(\lambda).
\ee
In the main text, we use the notion of an evaluation representation.
In the $t_{ij}^{(n)}$ basis, it is defined by the composition of the algebra homomorphism
\be
t_{ij}(\lambda) = \delta_{ij} + \frac{e_{ij}}{\lambda},
\ee
where $e_{ij}$ are the generators of $gl(N)$, with any
representation of $gl(N)$.  In particular, the Yangian generators
$t_{ij}^{(n)}$ with $n\geq1$ act as zero operators.

Alternatively, the Yangian \Ysl\ can also be defined as the unique
homogeneous quantization of $sl(N)[u]$ (the polynomial maps from the
complex plane to $sl(N)$)~\cite{Drin}.  It is generated by the two sets of elements $Q_0^a$ (a basis of $sl(N)$) and $Q_1^a$ subject to the following constraints
\beano
 &&[Q_0^a, Q_n^b] = i {f^{ab}}_c Q_n^c, \\
 &&{[Q_1^a,[Q_0^b,Q_1^c]]} + {[Q_1^b,[Q_0^c,Q_1^a]]} + {[Q_1^c,[Q_0^a,Q_1^b]]} =\\
 &&\qquad\qquad\qquad\qquad h^2 {f^a}_{pd}{f^b}_{qx}{f^c}_{ry}{f^{xy}}_e \kappa^{de}\ s_3(Q_0^p,Q_0^q,Q_0^r), \\
 &&{[[Q_1^a,Q_1^b],[Q_0^c,Q_1^d]]} + {[[Q_1^c,Q_1^d],[Q_0^a,Q_1^b]]} = \\
 &&\qquad h^2 \left( {f^a}_{pe}{f^b}_{qx}{f^{cd}}_y{f^y}_{rz}{f^{xz}}_g+ {f^c}_{pe}{f^d}_{qx}{f^{ab}}_y{f^y}_{rz}{f^{xz}}_g\right) \kappa^{eg}\ s_3(Q_0^p,Q_0^q,Q_1^r),
\eeano
where $\kappa^{ab}$ is the Killing form on $sl(N)$ and $s_3(.,.,.)$ is
the totally symmetrized product of three terms (normalized to $s_3(x,x,x)=x^3$).

The first presentation is helpful to identify the type of algebraic
structure generated by $\Tt(\lambda)$ in (\ref{RTtTt}).  The second
one is more convenient when constructing explicitly the Yangian
generators, as only two sets of elements are necessary to generate the whole Yangian \Ysl.

\sect{Construction of $J^a$}

To construct $J^a$, we heavily use the notation of internal spaces, which encodes both $su(N)$ indices and momenta.  Recall also that in this notation, $R_{ji} = R_{ij}^{-1}$.  The key observation in the construction of $J^a$ is that, due to the presence of $\delta_{ij}$ in (\ref{ZFa}), the commutator of $J^a_{(n)}$ with $a^\dag_0$ contains two contributions with different number of oscillators.  For the simplest case of $n=1$ one has
\be
[J^a_{(1)}, a^\dag_0] = a^\dag_0 a^\dag_{1}\left(R_{01} T^a_{1}R_{10}-T^a_{1}\right) a_{1}+ a^\dag_0 T^a_0.
\label{J0a*}
\ee
Comparing with (\ref{Ja*}), the term with no annihilation operator completely fixes $T^a_1 = t^a_1$.  In the commutator $[J^a,a^\dag_0]$, the only other term with one annihilation operator comes from the commutator $[J^a_{(2)},a^\dag_0]$ and we define $T^a_{12}$ so that these two contributions cancel.  Repeating this procedure for increasing number of oscillators uniquely determines the tensors $T^a_{1 \dots n}$.

In this Appendix, we adopt a different point of view, namely we prove that the solution given in (\ref{defT}) is indeed correct.

For generic $n$, one has
\be
[J^a_{(n)}, a^\dag_0] = a^\dag_0 a^\dag_{1\dots n}\left({\calr}_n^{-1}
T^a_{1\dots n}{\calr}_n-T^a_{1\dots n}\right) a_{n\dots 1}+ 
a^\dag_0 a^\dag_{1\dots n-1} B_n a_{n-1\dots 1},
\label{Jna*}
\ee
where
\bea
B_n &=& \sum_{i=1}^n{\calr}_{i-1}^{-1} T^a_{1\dots n|i}{\calr}_{i-1} \\
{\calr}_i &=&  R_{i0}\cdots R_{20}\, R_{10}\\
{\calr}_i^{-1} &=&  R_{01}\, R_{02}\cdots R_{0i}
\eea
Here the notation $T^a_{1\dots n|i}$ represents the indices substitutions $i\rightarrow 0$ and $k\rightarrow k-1$ for $k>i$.

Then, we simplify the expression for $B_n$:
\beano
 B_n &=& \sum_{i=1}^n\left( {\calr}_{i-1}^{-1}\left(\sum_{k=1}^{i-1} \alpha^n_k t^a_k +  \alpha^n_i t^a_0 + \sum_{k=i}^{n-1} \alpha^n_{k+1} t^a_k\right) {\calr}_{i-1}\right)\\
 &=& \sum_{i=1}^n\left( \sum_{k=1}^{i-1} \alpha^n_k {\calr}_{n-1}^{-1}t^a_k {\calr}_{n-1} +  \alpha^n_i {\calr}_{i-1}^{-1}t^a_0 {\calr}_{i-1} + \sum_{k=i}^{n-1} \alpha^n_{k+1} t^a_k \right)\\
 &=& {\calr}_{n-1}^{-1} \left(\sum_{0\leq k<i\leq n} \alpha^n_k t^a_k\right) {\calr}_{n-1} + \sum_{1\leq i\leq k\leq n-1} \alpha^n_{k+1} t^a_k + C_n,
\eeano
where $C_n$ is defined as
$$
 C_n = \sum_{i=1}^n \alpha^n_i {\calr}_{i-1}^{-1} t^a_0 {\calr}_{i-1}.
$$
Next, performing the independent sums on $i$ and using the properties
$$
\alpha_k^n = \frac{n-1}{n-k}\ \alpha_k^{n-1},
\qquad
\alpha_{k+1}^n = -\frac{n-1}{k}\  \alpha_k^{n-1},
$$
one gets
\beano
 B_n-C_n &=& {\calr}_{n-1}^{-1} \sum_{k=1}^{n-1} (n-k) \alpha^n_k
t^a_k {\calr}_{n-1} + \sum_{k=1}^{n-1} k\, \alpha^n_{k+1} t^a_k \\
 &=&(n-1){\calr}_{n-1}^{-1} \left(\sum_{k=1}^{n-1} \alpha^{n-1}_k
t^a_k\right) {\calr}_{n-1} - (n-1) \sum_{k=1}^{n-1} \alpha^{n-1}_{k+1}
t^a_k \nonumber\\
 &=& (n-1){\calr}_{n-1}^{-1} T^a_{1\dots n-1}{\calr}_{n-1}-
(n-1)T^a_{1\dots n-1}
\eeano
The simplification of $C_n$ is achieved using the $sl(N)$ invariance of the $R$-matrix, $[R_{0k},t^a_0+t^a_k]=0$, and the properties
$$
\sum_{i=1}^n \alpha^n_i = 0,
\qquad
\sum_{i=k+1}^n \alpha^n_i = - \alpha^{n-1}_k
$$
and leads to
\beano
C_n &=& \sum_{i=1}^n \alpha^n_i [{\calr}_{i-1}^{-1},t^a_0] {\calr}_{i-1}+ \sum_{i=1}^n \alpha^n_i t^a_0 \
 =\ \sum_{i=1}^n \alpha^n_i \sum_{k=1}^{i-1}{\calr}_{k-1}^{-1}[R_{0k},t^a_0] {\calr}_{k}\\
 &=& \sum_{k=1}^{n-1} \left(\sum_{i=k+1}^{n} \alpha^n_i \right) {\calr}_{k-1}^{-1}[R_{0k},t^a_0] {\calr}_{k}\
 =\ \sum_{k=1}^{n-1} \alpha^{n-1}_k {\calr}_{k-1}^{-1}[R_{0k},t^a_k] {\calr}_{k}\\
 &=&{\calr}_{n-1}^{-1}[T^a_{1\dots n-1},{\calr}_{n-1}] 
\eeano
Finally, we get that 
\be
B_n= (B_n-C_n)+C_n=n {\calr}_{n-1}^{-1}[T^a_{1\dots n-1},{\calr}_{n-1}]
\ee
so that the commutator (\ref{Jna*}) reduces to
$$
[J^a_{(n)}, a^\dag_0] = a^\dag_{01\dots n} {\calr}_{n}^{-1}[T^a_{1\dots n},{\calr}_{n}] a_{n\dots 1} - n a^\dag_{01\dots n-1} {\calr}_{n-1}^{-1}[T^a_{1\dots n-1},{\calr}_{n-1}] a_{n-1\dots 1}.
$$
Therefore, adjusting properly the coefficient of $J^a_{(n)}$ as in (\ref{defJ}), we obtain a complete cancellation of all the terms but one, yielding precisely the required commutation relation (\ref{Ja*}).  The proof of (\ref{Ja}) is similar.

\sect{Construction of $S^a$}
We shall show that $S^a$ is also of the form
$$
S^a = \sum_{n=1}^\infty \frac{(-)^{n+1}}{n!} a^\dag_{1\dots n} 
\wT^a_{1\dots n} a_{n\dots 1}
$$
and we shall determine the tensor $\wT^a_{1\dots n}$ directly from the commutation relations
\be
S^a = -\frac{i}{c_2} {f^{a}}_{bc} {[S^b, J^c]}= -\frac{i}{c_2} {f^{a}}_{bc}\sum_{n=1}^\infty \frac{(-)^{n+1}}{n!} {[S^b, J_{(n)}^c]},
\label{RRJ}
\ee
where ${f^{a}}_{bc}{f^{bc}}_d = c_2 \delta^a_d$.  We require $S^a$ to satisfy (\ref{Sa*}) and (\ref{Sa}) and consequently the right-hand side of (\ref{RRJ}) evaluates to
\beano
 {[S^b,J^c_{(n)}]} &=& [S^b, a^\dag_{1\dots n}] T^c_{1\dots n} a_{n\dots 1} + a^\dag_{1\dots n} T^c_{1\dots n} [S^b,a_{n\dots 1}]\\
 &=& \sum_{i=1}^{n} \left( a^\dag_{1\dots i-1} \, [S^b, a^\dag_i] \, a^\dag_{i+1\dots n} \, T^c_{1\dots n} \,  a_{n\dots 1} + a^\dag_{1\dots n} \,  T^c_{1\dots n} \, a_{n\dots i+1} \, [S^b, a_i] \, a_{i-1\dots 1}\right) \\
 &=& \sum_{i=1}^{n} \left( a^\dag_{1\dots i-1} \, \mu_i a^\dag_i t^b_i \, a^\dag_{i+1\dots n} \, T^c_{1\dots n} \,  a_{n\dots 1} - a^\dag_{1\dots n} \,  T^c_{1\dots n} \, a_{n\dots i+1} \, \mu_i t^b_i a_i \, a_{i-1\dots 1}\right) \\
 &+ & \frac{\rho g}{2} \sum_{i=1}^{n} {f^{b}}_{de} \, \left( a^\dag_{1\dots i-1} \, a^\dag_i t^{d}_i J^{e} \,  a^\dag_{i+1\dots n} T^c_{1\dots n} a_{n\dots 1} +a^\dag_{1\dots n} T^c_{1\dots n} a_{n\dots i+1} \, J^{d}t^{e}_i a_i \, a_{i-1\dots 1}\right)
\eeano
which shows that the solution we are looking for is
$$
S^a = S_I^a + S_{II}^a,
$$
where
\bea
 S_I^a &=& -\frac{i}{c_2} {f^{a}}_{bc} \sum_{n=1}^\infty \frac{(-)^{n+1}}{n!}
\, a^\dag_{1\dots n} \, \left[ \sum_{i=1}^{n} \mu_i t^b_i, T^c_{1\dots n} \right] \, a_{n\dots 1} \label{expS1} \\
 S_{II}^a &=& -\frac{i\rho g}{2 c_2} {f^{a}}_{bc} \sum_{n=1}^\infty \frac{(-)^{n+1}}{n!} \sum_{i=1}^{n} {f^{b}}_{de} \, \Big( a^\dag_{1\dots i} \, t^{d}_i J^{e} \,  a^\dag_{i+1\dots n} T^c_{1\dots n} a_{n\dots 1} \nonumber \\
 &&+a^\dag_{1\dots n} T^c_{1\dots n} a_{n\dots i+1}J^{d}t^{e}_i a_{i\dots 1}\Big) .\nonumber
\eea

These expressions can be considerably simplified.  In $S_I^a$ we introduce the known form of $T^c_{1\dots n} = \sum_{k=1}^n \, \alpha^n_k \, t^c_k$ and with the definition of the second Casimir we get
$$
i{f^{a}}_{bc}\left[ \sum_{i=1}^{n} \mu_i t^b_i, T^c_{1\dots n} \right]=
-{f^{a}}_{bc} \sum_{i=1}^{n} {f^{bc}}_e \mu_i \alpha^n_i t^e_i =
-c_2 \sum_{i=1}^{n} \mu_i \alpha^n_i t^a_i
$$
so that
\be
S_I^a = \sum_{n=1}^\infty \frac{(-)^{n+1}}{n!}
\, a^\dag_{1\dots n} \, \left(  \sum_{i=1}^{n} \mu_i \alpha^n_i t^a_i \right) \, a_{n\dots 1} .
\ee

The next step is to simplify the contribution $S_{II}^a$.  We shift $J^e,J^d$ towards $T^c$ using
\beano
{[J^e,a^\dag_{i+1\dots n}]} &=& a^\dag_{i+1\dots n} (\sum_{j=i+1}^n t_j^e) \\
{[J^d,a_{n\dots i+1}]} &=&- (\sum_{j=i+1}^n t_j^d) a_{n\dots i+1}
\eeano
and with the anti-symmetry of ${f^b}_{de}$ we get
\beano
S_{II}^a &=& -\frac{i\rho g}{2 c_2} {f^{a}}_{bc}
\sum_{n=1}^\infty \frac{(-)^{n+1}}{n!} 
\sum_{i=1}^{n} {f^{b}}_{de} \, \left( 
a^\dag_{1\dots n} \,\left[ \sum_{j=i+1}^n t_i^d t_j^e, T^c_{1\dots n}
\right] a_{n\dots 1} \right.\\
 &&+ \left. \vphantom{\left[ \sum_{j=i+1}^n\right]}
a^\dag_{1\dots n} \,\left[ t_i^d , T^c_{1\dots n}
\right] J^e a_{n\dots 1}
\right).
\eeano
The second term is simplified using the commutator
$$
\sum_{i=1}^{n} [ t_i^d , T^c_{1\dots n} ] = i{f^{dc}}_g \, T^g_{1\dots n}
$$
and the identity 
$$
{f^a}_{bc}{f^b}_{de}{f^{dc}}_{g} =  -\frac{c_2}{2}\,{f^a}_{eg} .
$$
That same identity, combined with the explicit expression for $T^c_{1\dots n}$, helps to reduce the first term to
$$ 
{f^a}_{bc}{f^b}_{de} \left[ \sum_{1\leq i<j \leq n} t_i^d t_j^e, \sum_{k=1}^n \, \alpha^n_k  t^c_k \right] =
i \frac{c_2}{2}{f^a}_{bc}\, \sum_{1\leq i<j \leq n} (\alpha^n_i+\alpha^n_j)t_i^b t_j^c .
$$
Altogether, the expression we find for $S_{II}^a$ is
\be
S_{II}^a = -\frac{ \rho g}{4} {f^a}_{bc} \sum_{n=1}^\infty \frac{(-)^n}{n!} a^\dag_{1\dots n} \,\left(\sum_{1\leq i<j \leq n} (\alpha^n_i+\alpha^n_j)t_i^b t_j^c
\ +\ T^b_{1\dots n} J^c\right)\, a_{n\dots 1} .
\label{finalR}
\ee

We can merge the two contributions to (\ref{finalR}).  We plug in the expansion (\ref{defJ}) for $J^c$ in the second one and appropriately label the auxiliary spaces
\beano
\lefteqn{ \sum_{k=1}^\infty  \frac{(-)^{k}}{k!}
\, {f^a}_{bc}a^\dag_{1\dots k} \,T^b_{1\dots k} J^c\, 
a_{k\dots 1}} \\
 &= & {f^a}_{bc}\sum_{k=1}^\infty \sum_{l=1}^\infty
\frac{(-)^{k+l+1}}{k! l!} a^\dag_{1\dots k} a^\dag_{k+1\dots k+l}
\, T^b_{1\dots k} T^c_{k+1\dots k+l} \, a_{k+l\dots k+1} a_{k\dots 1} \\
 &= & {f^a}_{bc}\sum_{n=2}^\infty \frac{(-)^{n+1}}{n!}
a^\dag_{1\dots n} \left( \sum_{k=1}^{n-1} \left({n \atop k} \right) T^b_{1\dots k} T^c_{k+1\dots n}\right) a_{n\dots 1}
\eeano
so that we get for $S_{II}^a$
\be
S_{II}^a = -\frac{\rho g}{2} {f^a}_{bc}\sum_{n=2}^\infty \frac{(-)^{n+1}}{n!}
a^\dag_{1\dots n} \, T^{bc}_{1\dots n} \, a_{n\dots 1} ,
\ee
where the new tensor $T^{bc}_{1\dots n}$ is defined below and turns out to be surprisingly simple
\beano
T^{bc}_{1\dots n} &=& \frac{1}{2} \left(\sum_{k=1}^{n-1} \left({n \atop k}
\right) T^b_{1\dots k} \, T^c_{k+1\dots n} - \sum_{1 \leq i<j \leq n}
(\alpha^n_i+\alpha^n_j)\, t_i^b t_j^c \right) \\
 &=& \frac{1}{2} \sum_{1 \leq i<j \leq n} \left( \sum_{k=i}^{j-1} \left(
{n \atop k} \right) \alpha^k_i \alpha^{n-k}_{j-k} - \alpha^n_i - \alpha^n_j
\right) \, t_i^b t_j^c \\
 &=& - \sum_{1 \leq i<j \leq n} \alpha^n_j \, t_i^b t_j^c .
\eeano
In the last step, we have used the property 
(proved in the next section)
\be
\sum_{k=i}^{j-1} 
\bin{n}{k}
\alpha_i^k \alpha_{j-k}^{n-k} 
- \alpha_i^n = -\alpha_j^n \label{proprioA} .
\ee
Putting together $S_{I}^a$ and $S_{II}^a$, we get the final expression for $S^a$
\be
S^a = \sum_{n=1}^\infty \frac{(-)^{n+1}}{n!} a^\dag_{1\dots n} \left(
\sum_{i=1}^{n} \mu_i \alpha^n_i t^a_i +\frac{\rho g}{2} {f^a}_{bc}\sum_{1 \leq i<j \leq n}
\alpha^n_j \, t_i^b t_j^c
\right) a_{n\dots 1} .
\ee
The first few terms of this series are
\beano
 S^a_{1} &=& a^\dag_{1}\, \mu_1 t^a_1 \,  a_{1}, \\
 S^a_{2} &=& -\frac{1}{2} a^\dag_{12}\, \left( \mu_1 t^a_1 - \mu_2 t^a_2
- \frac{\rho g}{2} {f^{a}}_{bc} t^b_1 t^c_2\right) \,  a_{21}, \\
 S^a_{3} &=& \frac{1}{6} a^\dag_{123}   \left( \mu_1 t^a_1 - 2 \mu_2 t^a_2 +
\mu_3 t^a_3 - \frac{\rho g}{2} {f^a}_{bc}(2t_1^b t_2^c - t_1^b t_3^c
-t_2^b t_3^c) \right)a_{321}.
\eeano
Recall that there is an implied integration on $\mu_i$ in these expressions.

\subsection{Proof of the property (\ref{proprioA})}

We want to show that for $1 \leq i < j \leq n$ we have
$$
\sum_{k=i}^{j-1} \bin{n}{k} \alpha_i^k \alpha_{j-k}^{n-k} - \alpha_i^n = 
-\alpha_j^n.
$$
It is equivalent to show that $f(m)=g(m)$ for $m$ integer, where
\beano
f(m) &=& \sum_{k=0}^{m-1} \bin{n}{k+i} \alpha_i^{k+i} \alpha_{m-k}^{n-k-i}
- \alpha_i^n, \\
g(m) &=& -\alpha_{i+m}^n.
\eeano
This is done by recursion. Obviously $f(1)=g(1)$ and we then show that $df(m)\equiv f(m+1)-f(m)$ and $dg(m)\equiv g(m+1)-g(m)$ are equal.
$$
dg(m)= (-)^{i+m+1} \bin{n}{i+m}
$$
while
\beano
df(m) &=& \sum_{k=0}^{m} \bin{n}{k+i} \alpha_i^{k+i} \alpha_{m-k+1}^{n-k-i} - \sum_{k=0}^{m-1} \bin{n}{k+i} \alpha_i^{k+i} \alpha_{m-k}^{n-k-i} \\
 &=& \sum_{k=0}^{m-1} (-)^{i+m+k+1}\bin{n}{k+i}\bin{k+i-1}{i-1} \left\{
\bin{n-k-i-1}{m-k} \right. \\
 &&+ \left. \bin{n-k-i-1}{m-k-1} \right\}+(-)^{i+1}\bin{n}{m+i}
\bin{m+i-1}{i-1} \\
 &=& \sum_{k=0}^{m-1} (-)^{i+m+k+1}\bin{n}{k+i}\bin{k+i-1}{i-1} 
\bin{n-k-i}{m-k} \\
 && + (-)^{i+1}\bin{n}{m+i} \bin{m+i-1}{i-1} \\
 &=&(-)^{i+m+1}\bin{n}{m+i} \left\{{(i+m)! \over (i-1)! m!} \sum_{k=0}^{m-1}
{(-)^k \over k+i} \bin{m}{k} \right. \\
&& + \left. (-)^m \bin{m+i-1}{i-1} \right\} \\
 &=& (-)^{i+m+1}\bin{n}{m+i} \left\{ {(i+m)! \over (i-1)!\, m!}  \sum_{k=0}^{m} {(-)^k \over k+i} \bin{m}{k} \right\}\\
&\equiv& dg(m)\ \left\{ h(m) \right\}.
\eeano
We need to show that $h(m)=1$, which is implied by
\beano
\sum_{k=0}^m \frac{(-)^k}{k+i} \bin{m}{k} &= &\sum_{k=0}^m \int_0^1\ dx\ (-)^k
x^{k+i-1} \bin{m}{k}
= \int_0^1\ dx\ x^{i-1} \sum_{k=0}^m \bin{m}{k} x^{k} \\
&=&
\int_0^1\ dx\ x^{i-1} (1-x)^m 
=  \frac{\Gamma(i)
\Gamma(m+1)}{\Gamma(m+i+1)}.
\eeano
This ends the proof.

%
%


\begin{thebibliography}{99}

\bibitem{BGHP} F.D.~Haldane, Z.N.~Ha, J.C.~Talstra, D.~Bernard, V.~Pasquier, Phys. Rev. Lett {\bf 69} (1992) 2021;\\
D.~Bernard, M.~Gaudin, F.D.~Haldane, V.~Pasquier, J. Phys. A:Math. Gen. {\bf 26} (1993) 5219.

\bibitem{ABB} J. Avan, O. Babelon, E. Billey, Phys. Lett. {\bf A 118} (1994) 263.

\bibitem{Wad} S. Murakami, M. Wadati, J. Phys. {\bf A29} (1996)
  7903.

\bibitem{ZF} A.B. Zamolodchikov, A.B. Zamolodchikov, Ann. Phys. {\bf
    120} (1979) 253; L.D. Faddeev, Sov. Scient. Rev. {\bf C1} (1980) 107.

\bibitem{PWZ} F. Pu, Y. Wu, B. Zhao, J. Phys. A: Math. Gen. {\bf 20} (1987) 1173.

\bibitem{Gut} E. Gutkin, Phys. Rep. {\bf 167} (1988) 1.

\bibitem{fock} A. Liguori, M. Mintchev, Comm. Math. Phys. 
{\bf 169} (1995) 635; Lett. Math. Phys. {\bf 33} (1995) 283;\\
A. Liguori, M. Mintchev, M. Rossi, J. Math. Phys. 
{\bf 38} (1997) 2888.

\bibitem{Ros} R. Rosales, Stud. Appl. Math. {\bf 59} (1978) 117.

\bibitem{solQ}  E. Sklyanin, L. D. Faddeev, Sov. Phys. Dokl.
   {\bf 23} (1978) 902; \\
E. Sklyanin, Sov. Phys. Dokl. {\bf 24} (1979) 107;\\
H.B. Tacker, D. Wilkinson, Phys. Rev. {\bf D19} (1979) 3660;\\
 D.B. Creamer, H.B. Tacker, D. Wilkinson, Phys. Rev. {\bf 
D21} (1980) 1523;\\
 J. Honerkamp, P. Weber, A. Wiesler, Nucl. Phys. {\bf
B152} (1979) 266;\\
 B. Davies, J. Phys. {\bf A14} (1981) 2631.

\bibitem{CP} V. Chari, A. Pressley, {\it A guide to quantum groups},
  chap. 12, Cambridge University Press, 1994.

\bibitem{13} E. Ragoucy, P. Sorba, {\em Yangian realizations from
    finite $W$-algebras}, ENSLAPP-AL-672/97, {\tt hep-th/9803243}, to appear 
in Comm. Math. Phys;
Cz. J. Phys. {\bf 48} (1998) to appear, {\tt hep-th/9803242}.

\bibitem{BR} C. Briot, E. Ragoucy, in preparation.

\bibitem{BRS} F. Barbarin, E. Ragoucy, P. Sorba, Czech. J. Phys. {\bf 46} (1996) 1165,
{\tt hep-th/9612070}.

\bibitem{Aris} A. E. Aristein, Phys. Lett. {\bf B95} (1980) 280;\\
F. A. Smirnov, {\sl Form factors in completely integrable models of
  QFT}, World Scientific, 1992.

\bibitem{GLM} M. Gattobigio, A. Liguori, M. Mintchev,
Phys. Lett. {\bf B428} (1998) 143; {\tt hep-th/9811188} to appear in
J. Math. Phys.\\
A. Liguori, M. Mintchev, L. Zhao, Comm. Math. Phys. 
{\bf 194} (1998) 569;

\bibitem{Bound} G. Olshanski, Lect. Note Math. {\bf 1510}, Springer,
  Berlin (1992), pp.103-120;\\
 E.K. Sklyanin, J. Phys. {\bf A21} (1988) 2375;\\
 I.V. Cherednik, Theor. Math. Phys. {\bf 61} (1984) 977.

\bibitem{Molev} A. Molev, M. Nazarov, G. Olshanski, {\tt hep-th/9409025}.

\bibitem{MRSZ2} M. Mintchev, E. Ragoucy, P. Sorba, Ph. Zaugg, in preparation.

\bibitem{Drin} V.G. Drinfel'd, Sov. Math. Dokl. {\bf 32} (1985) 254.

\end{thebibliography}
\end{document}